\documentclass[journal,twoside]{IEEEtran}
\IEEEoverridecommandlockouts                              

\usepackage{graphics} 
\usepackage{epsfig} 
\usepackage{amsmath} 
\usepackage{amssymb}  
\usepackage{cite}
\usepackage{xcolor}
\usepackage{eurosym}
\usepackage{bm}
\usepackage{empheq}
\usepackage{float}
\usepackage{caption}
\usepackage{subcaption}
\usepackage{chngcntr}

\usepackage{enumerate}
\usepackage{hyperref}
\usepackage{multicol}
\usepackage{cuted}
\usepackage{xcolor,cite,etoolbox}
\makeatletter 
\pretocmd\@bibitem{\color{black}\csname keycolor#1\endcsname}{}{\fail}
\newcommand\citecolor[1]{\@namedef{keycolor#1}{\color{blue}}}
\makeatother

\allowdisplaybreaks

\newcommand{\norm}[1]{\lVert#1\rVert}
\newcommand{\abs}[1]{\lvert#1\rvert}

\newtheorem{theorem}{Theorem}

\newtheorem{assumption}{Assumption}

\title{%
Hierarchical Power Flow Control in Smart Grids: Enhancing Rotor Angle and Frequency Stability with Demand-Side Flexibility
\thanks{This work was supported by ARPA-E Award DE-AR0000702, with additional support from the Finite Earth Initiative (funded by Leslie and Mac McQuown).}}
\author{
Chao~Duan,
Pratyush~Chakraborty,
Takashi~Nishikawa,
and~Adilson E.~Motter%
\vspace{-0.3cm}
\thanks{Chao Duan, Takashi Nishikawa, and Adilson E. Motter are with the Department of Physics and Astronomy, Northwestern University, Evanston, IL 60208, USA (e-mail: chao.duan@northwestern.edu; t-nishikawa@northwestern.edu; motter@northwestern.edu).}%
\thanks{Pratyush Chakraborty was with the Department of Physics and Astronomy, Northwestern University, Evanston, IL 60208, USA. He is now with the Department of Electrical and Electronics Engineering, Birla Institute of Technology \& Science (BITS), Pilani, 333031, India (e-mail: pchakraborty@hyderabad.bits-pilani.ac.in).}
}

\begin{document}

\maketitle

\begin{abstract}
Large-scale integration of renewables in power systems gives rise to new challenges for keeping synchronization and frequency stability in volatile and uncertain power flow states. To ensure the safety of operation, the system must maintain adequate disturbance rejection capability at the time scales of both rotor angle and system frequency dynamics. {This calls for flexibility to be exploited on both the generation and demand sides, compensating volatility and ensuring stability at the two separate time scales. This article proposes a hierarchical power flow control architecture that involves both transmission and distribution networks as well as individual buildings to enhance both small-signal rotor angle stability and frequency stability of the transmission network.} The proposed architecture consists of a transmission-level optimizer enhancing system damping ratios, a distribution-level controller following transmission commands and providing frequency support, and a building-level scheduler accounting for quality of service and following the distribution-level targets. We validate the feasibility and performance of the whole control architecture through real-time hardware-in-loop tests involving real-world transmission and distribution network models along with real devices at the Stone Edge Farm Microgrid. 
\end{abstract}



\section{Introduction}
\label{sec:intro}
\IEEEPARstart{F}LEXIBILITY is the most valuable asset in modern power systems with heavy penetration of renewable energy. Due to the volatility and uncertainty of renewable sources, it is the system's flexibility that determines how much renewable energy can be integrated. Conventionally, system flexibility comes only from the generation side, where the excitation system is carefully designed to keep the synchronization among the generators while the generator mechanical power is controlled by the turbine-governor system to maintain the real-time supply-demand balance at each node of the network \cite{machowski2011,wood2013}. {To optimally exploit generation-side flexibility, researchers have extensively investigated and implemented economic dispatch \cite{Lin2017Power}, optimal power flow (OPF) \cite{CapitanescuCritical}, automatic generation control (AGC) \cite{Jaleeli1992Understanding}, and {stability-constrained OPF \cite{867137,jiang2010reduced,pizano2011new}}.} This {generation-oriented control} framework worked well for the entire 20th century---a period over which intermittent power sources played a minor role in the system. Yet, {in addition to the generation side, there is a significant amount of untapped flexibility that can be exploited from the demand side to support the operation of current and future renewable-integrated power systems.}

{The demand-side management (DSM) is a portfolio of measures and policies that uses load as an additional degree of freedom to make the power network more secure, efficient, and environmentally friendly \cite{strbac2008demand,palensky2011demand,samadi2012advanced}. Since the emergence of the concept of DSM in the 1980s \cite{gellings1985concept}, its benefits have been appreciated at different time scales of power system operation.} The majority of DSM programs are focused on the time scale of one hour or longer to shift the steady-state load pattern, which alleviates the system balancing challenges and improves the utilization of renewables; in return, customers obtain rebates or other compensations without necessarily reducing their total usage of electricity \cite{bao2018optimal,aghaei2013demand}. At the time scale of minutes, on the other hand, DSM is also considered a source of regulating ancillary service to counter sub 5-min volatility and forecast errors \cite{aghaei2013demand}. Despite requiring a faster response of the load, the problem at this time scale still concerns only steady states and thus involves only the algebraic power flow balancing equations. {To actually realize the idea of DSM at this time scale, a massive number of distributed loads need to be integrated, controlled, and coordinated to create grid-scale effects. This can be achieved using buildings-to-grid (BtG) integration \cite{mirakhorli2018model, razmara2017building, liu2018coordinating,taha2017buildings,rey2018strengthening,dong2018occupancy}, in which market mechanisms and control strategies are developed to incentivize participation in DSM programs.}

Researchers have also recognized the potential of DSM to act at the time scale of seconds and participate in frequency regulation \cite{teng2015benefits}. This possibility becomes realistic when the system is equipped with fast responding devices, especially energy storage ones. The idea of load participation in frequency control was conceived in the late 1970s. In \cite{schweppe1980}, the authors advocated for a large-scale deployment of loads that could assist or even replace turbine governor systems. They proposed the use of spot prices to incentivize users to modify their consumption accordingly. They also emphasized the fact that such frequency-responsive loads would allow the system to accept more fluctuating energy sources, such as wind and solar generation. Real-time DSM based on frequency has been modeled and simulated in various studies \cite{short2007,molina2011,moghadam2014, meyn2015}. In \cite{zhao2014,mallada2017,Shiltz2017}, load side primary and secondary frequency control was developed with stability and optimality considerations. The literature has shown that many different appliances (such as washing machines, refrigerators, and heating, ventilation, and air conditioning units) and plug-in hybrid electric vehicles, though small in individual capacities, can be controlled collectively and coordinated to help regulate system frequency. {A BtG framework aiming at regulating system frequency was discussed in \cite{taha2017buildings,rey2018strengthening,dong2018occupancy}.}

{Another aspect of power system stability problem is the small-signal rotor angle stability, which concerns the synchronization of generators under small disturbances. The existing attempts to enhance this stability can be generally classified into two groups. One group focuses on the design of power system stabilizers (PSS) or other control parameters for excitation systems and flexible AC transmission system (FACTS) devices \cite{aboul1996damping,borsche2015effects,konara2015robust,CaiSimultaneous,pal2006robust}. The damping controllers work on the time scale of power system electrical-mechanical dynamics and provide auxiliary signals for actuators (exciters or FACTS devices) to damp system oscillations. However, damping controllers are usually designed for merely one or a few selected operation points of the system and may not work well if unexpected operation scenarios are encountered in practice, especially in a system with high penetration of intermittent renewables. {The potential of controlling load to improve rotor angle stability was explored in \cite{lian2018wide,wilches2019damping,zhang2018investigations}. These studies proposed to use fast load response (at sub-second time scale) to damp electromechanical oscillations of the system. From our field test experience, these methods are unrealistic and impossible to implement in practice because most loads have a much slower response (at the time scale of seconds to minutes) and time delays involved in the control loop are significant (at the time scale of seconds), especially when a large number of loads need to be controlled.} The other group of attempts to enhance small-signal rotor angle stability focuses on changing the operating points of the system by re-dispatching generation units \cite{chung2004generation,zarate2010opf,li2013eigenvalue,li2016sqp,li2019sequential}. This naturally leads to the formulation of small-signal stability constrained optimal power flow (SSSC-OPF) \cite{li2013eigenvalue,li2016sqp,li2019sequential}, in which constraints for system damping ratios are added to a conventional OPF problem. The control decision is thus implemented at every dispatch interval (normally ranging from 5 to 15 min) to make sure the system operates at a well-damped equilibrium. {Empirical studies \cite{tang2017assessment, hu2011small} have revealed that the control of flexible load can also alter the power flow and drive the system toward more stable operation points. To the best of the authors' knowledge, none of the existing research proposes a systematic problem formulation and algorithm to \textit{co-optimize} generation and demand to enhance small-signal rotor-angle stability.} We show in this work that the load-side flexibility can be exploited to substantially improve the small-signal rotor angle stability by coordinated power flow optimization together with generation-side response. The idea is in line with the SSSC-OPF methods \cite{zarate2010opf,li2016sqp}, but we emphasize the fact that our control approach coordinates adjustments of not only generation but also load to effectively drive the system towards a more stable operation equilibrium. }

{Hierarchical control structures have been previously designed for electricity market operation \cite{manshadi2015hierarchical}, management of integrated community energy systems \cite{xu2015hierarchical}, demand response at different time scales \cite{bhattarai2016design}, and coordination between transmission and distribution operators \cite{yuan2017hierarchical}. 
This article develops and implements a three-level, hierarchically coordinated, generation and demand control that can simultaneously enhance small-signal rotor angle and frequency stability of the transmission system. Our principal contributions are as follows:
\begin{enumerate}[---]
    \item \emph{{Hierarchical exploitation of demand-side flexibility to enhance transmission network stability}}. {The proposed coordination of control algorithms in different layers {enhances both small-signal rotor angle stability and frequency stability} of the transmission network under a unified framework, while respecting} various security constraints relevant to the transmission network, distribution networks, and appliances in the buildings. {The hierarchical control incorporates small-scale building-level loads while preserving the numerical scalability of the problem.} 
    \item \emph{Generation and demand co-optimization method for stability enhancement}. We propose a generation and demand coordinated OPF method that seeks the best generator re-dispatch and demand response strategy to maximize the system's minimal damping ratio. 
    The problem is formulated based on the power system nonlinear differential-algebraic equation (DAE) model with analytic derivative information for system eigenvalues. Piece-wise linear approximation and trust-region methods are integrated to handle the non-smoothness and non-linearity of the problem, leading to a sequential nonlinear optimization algorithm. The method is applied to a real-world transmission system with 7459 buses.
    \item \emph{Integrated hardware-in-the-Loop (HiL) implementation}. We design and implement a real-time HiL test of the proposed control architecture. The prominent features of this test include the real-time co-simulation of large-scale transmission and distribution systems, the integration of three-level optimization and control algorithms with the real-time simulation, and the interconnection of the simulation with a real microgrid. With this implementation, the responses of real building-level loads and their impact on the system-level dynamics are well reflected in the test.
\end{enumerate}}


\vspace{0cm}
\section{Hierarchical Control Structure}
\label{sec:Hcon}

\begin{figure}[t]
	\centering
	\includegraphics[width=3.5in]{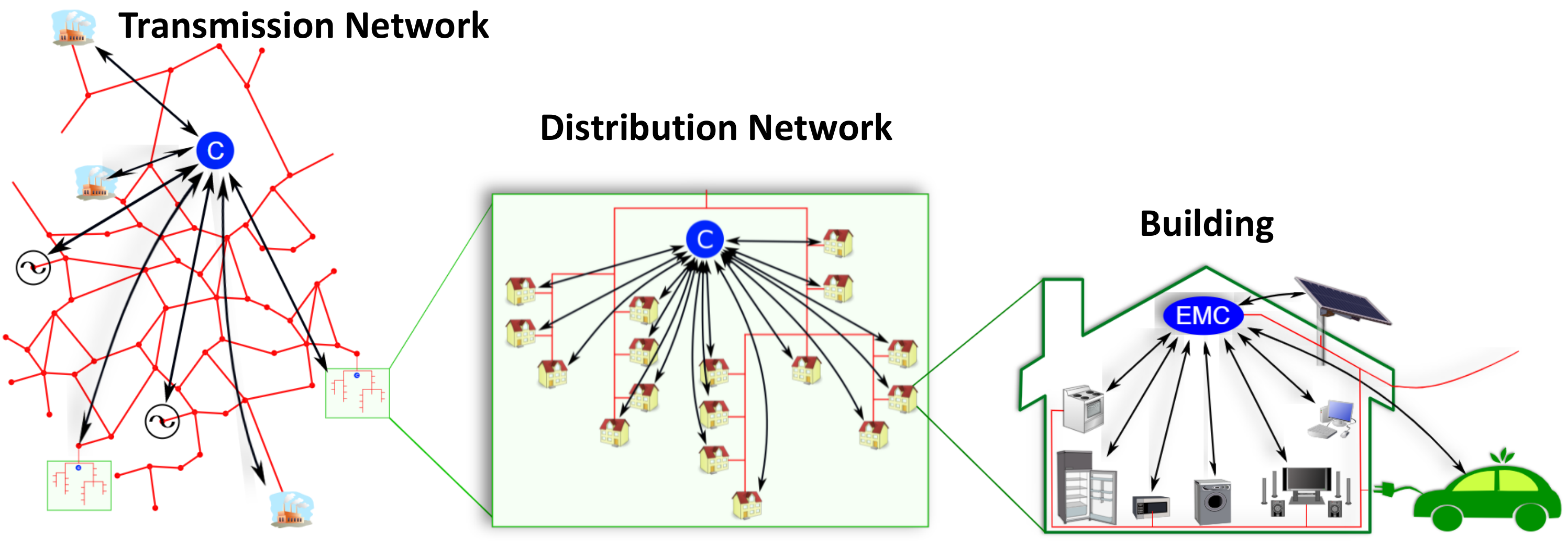}
	\caption{Hierarchical control structure. The proposed approach operates hierarchically on the transmission network (left), distribution networks (center), and individual buildings (right).} \label{controlstructure}
\end{figure}

\begin{figure}[t]
	\centering
	\includegraphics[width=3.2in]{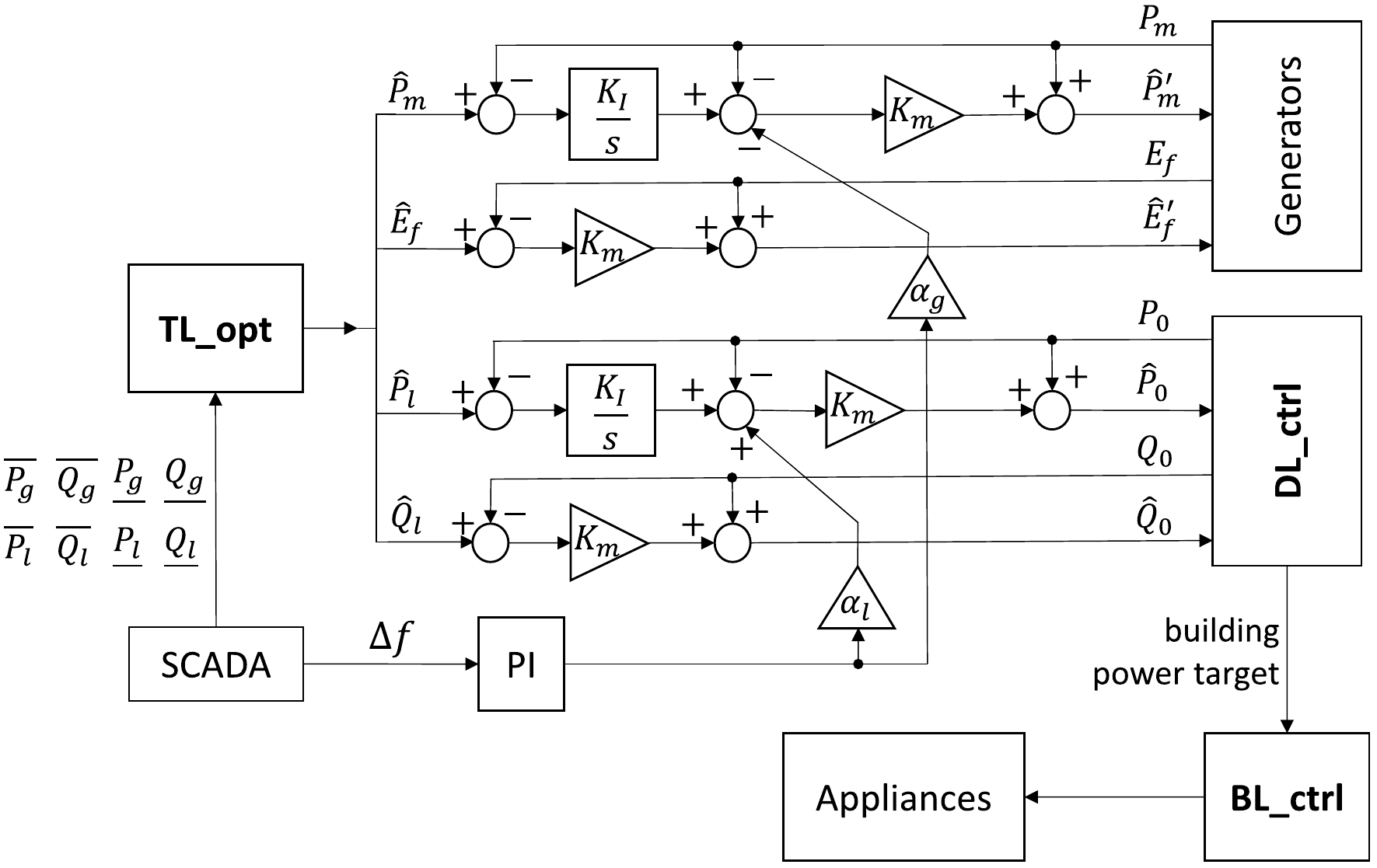}
	\caption{{Hierarchical control diagram, where TL\_opt is the transmission-level optimization, DL\_ctrl is the distribution-level control, and BL\_ctrl is the building-level controller.}}  \label{ctrldiagram}
\end{figure}

{To enhance both rotor angle and frequency stability within a single framework, the three-level controllers shown in Fig. \ref{controlstructure}, with details given in Fig.~\ref{ctrldiagram}, are designed for the transmission, distribution, and building levels. 
{At the beginning of every optimization cycle (of length 1 min in our implementation), the transmission-level optimizer TL\_opt checks whether the system requires small-signal stability enhancement and whether the previous optimization decisions have been successfully implemented. If that is the case, the transmission-level optimization algorithm is executed to compute setpoints for generator mechanical powers, generator exciter voltages, load active powers, and load reactive powers given the current state and available flexibility of the system. The power targets are then sent to the generators and also to the distribution networks.
}
Then, the distribution-level feedback controller DL\_ctrl computes the power target for each building in real-time to eliminate the control error at the point of common coupling (PCC). The building-level optimization BL\_opt is invoked every {1 s} to schedule the controllable appliances to follow the power target received from DL\_ctrl. The generators respond to system frequency deviations through the conventional AGC mechanism. In addition to that, the distribution networks also directly respond to system frequency deviations through a parallel proportional controller.
}

\section{Transmission-level Problem and Algorithm}
\label{sec:Tcon}
\subsection{Transmission System Dynamical Model}
{On the time scale of small-signal stability problems for rotor angles, the generator mechanical power and load demand are treated as constants.  However, their values can influence the operation point of the system and thus the small-signal stability.  It is this influence that allows demand control to impact the small-signal stability.} We employ the two-axis model
\cite{machowski2011,sauer2017power} to represent the generators. In a compact form, the system differential equations are written as
\begin{subequations}
\begin{flalign}
   & \dot{{\delta}} = {\omega}-{\omega}_s, \label{D1}\\
    &  \dot{{\omega}} = M^{-1}\big({p}_m-\text{Re}\{[C_gV]Y_g^*V^*\}-D({w}-{\omega}_s)\big),\label{D2}\\
    &   \dot{{e}}'_q =T'^{-1}_{d0} \big( {e}_f - {e}'_q - ({X}_d-{X}'_d) \text{Re}\{[\text{e}^{j(\frac{\pi}{2}-{\delta})}]Y_gV\}   \big),\label{D3}\\
    &    \dot{{e}}'_d =T'^{-1}_{q0} \big(  - {e}'_d + ({X}_q-{X}'_q) \text{Im}\{[\text{e}^{j(\frac{\pi}{2}-{\delta})}]Y_gV\}  \big),\label{D4}
\end{flalign}
\end{subequations}
and the algebraic equations are given by
\begin{subequations}
\begin{flalign}
   &   {e}'_d = \text{Re}\{[\text{e}^{j(\frac{\pi}{2}-{\delta})}](C_g+jX_q'Y_g)V \}, \label{A1}\\
    &  {e}'_q = \text{Im}\{[\text{e}^{j(\frac{\pi}{2}-{\delta})}](C_g+jX_d'Y_g)V \},\label{A2}\\
    &   {P}_l + \text{Re}\{ [C_l V]Y_l^* V^* \} = 0,\label{A3}\\
    &    {Q}_l + \text{Im}\{ [C_l V]Y_l^* V^* \} = 0.\label{A4}
\end{flalign}
\end{subequations}
Here, the square bracket $[{x}]$ represents the diagonal matrix constructed from vector ${x}$, and $x^*$ denotes the complex conjugate of $x$. The {time-dependent} state variables ${\delta}$, ${\omega}$, ${e}'_q$, and ${e}'_d$ are the vector of generator power angles, electrical angular speeds, q-axis transient voltages, and d-axis transient voltages, respectively. The ${\omega}_s$ is the nominal system frequency in rad/s. The constants $M$, $D$, $T'_{q0}(T'_{d0})$, $X_q(X_d)$, and $X'_q(X'_d)$ are the diagonal matrices of generator inertia, damping coefficients, q-axis (d-axis) transient time constants, q-axis (d-axis) synchronous reactances, and q-axis (d-axis) transient reactances, respectively. Since the damper windings are modeled in the two axis model, $D$ only represents the effects of mechanical friction, which is almost always negligible \cite{Lima2015Benchmark}. In addition, to represent the power flow equation of the system, we adopt the MATPOWER style notations \cite{zimmerman2010matpower}. $V$ is the {time-dependent} vector of nodal complex voltages. $C_g$ is the $n_g\times n_b$ mapping matrix from all buses to generator buses. $Y$ is the system admittance matrix and $Y_g=C_gY$. Similarly, $C_l$ is the mapping from all buses to load buses and $Y_l=C_lY$. The variables ${p}_m$ and ${e}_f$ are the vectors of generator mechanical power and exciter voltages. Finally, $P_l$ and $Q_l$ are the vectors of active and reactive loads.

The exciter is the major control point in power systems for achieving voltage regulation and synchronization enhancement. In this article, the transmission-level control aims to reschedule the power flow under a fixed configuration of the excitation system to improve damping performance. Therefore, we need a way to represent the effects of the excitation system and power system stabilizer (PSS). In theory, we can incorporate the detailed models of exciters and PSSs. However, doing so would at least double the order of the model, which would be computationally disadvantageous for real-time online applications on large-scale systems. The time constants of exciters are usually small and PSSs are designed to provide phase shifts so that the electrical torques are approximately in phase with rotor speeds. Hence, we adopt the following simplified equation as a coarse model for exciters and PSSs:
\begin{equation}
    {e}_f = E_f -K_AC_g(\abs{V}-\abs{V_{e}}) + K_AK_S({\omega}-{\omega}_s),
\end{equation}
where $E_f$ is the steady-state exciter voltage, and $K_A$ and $K_S$ are the diagonal matrices whose nonzero elements are the gains of the exciters and PSSs. Both values of the gains are chosen as the open-loop gains at the local mode frequency, which we set to be 1.5Hz. $V_{e}$ is the complex voltage for the targeted power flow equilibrium. 

Given a power flow solution $V_e=[{\mathcal{V}}_e]\text{e}^{j{\theta}_e}$, we can linearize the above nonlinear dynamics around this equilibrium and obtain a linear DAE of the form
\begin{equation}\label{DAEcomp}
    \begin{pmatrix}
    I & {0}\\
    {0} & {0}
    \end{pmatrix}
    \begin{pmatrix}
    \dot{{s}} \\ \dot{{x}}
    \end{pmatrix}
    =
    \begin{pmatrix}
    A({V}_e) & B({V}_e)\\
    C({V}_e) & D({V}_e)
    \end{pmatrix}
    \begin{pmatrix}
    {s}\\
    {x}
    \end{pmatrix},
\end{equation}
where ${s}=({\delta},{\omega},{{e}}'_q,{{e}}'_d)$ and ${x}=({\theta},{\mathcal{V}})$. {Details on the formulation of the coefficient matrices are included in the the technical notes of our GitHub repository\footnote{\label{note1}See \href{https://github.com/cduan2020/EigenOPF/blob/master/technical_notes.pdf}{https://github.com/cduan2020/EigenOPF/blob/master/technical\_notes.pdf} within our GitHub repository \cite{GitHubref}.}.} Note that the linearized system matrices are purely determined by the power flow solution $V_e$ (equivalently, by ${x}_e$). For notational simplicity, we drop the subscript ``${e}$'' for the rest of this article.

The generalized eigenvalues of this linearized system determines the small-signal stability of the system at the operation point. In practice, $D(V)$ is always invertible. Therefore, the generalized eigenvalue problem reduces to a standard eigenvalue problem of a much smaller matrix:
\begin{equation}
    J\left(V\right)=A\left(V\right)-B(V){D\left(V\right)}^{-1}C(V).
\end{equation}
The damping ratios ${\xi}_i=-\text{Re}(\lambda_i)/\abs{\lambda_i}$
of eigenvalues $\lambda_i$ of $J\left(V\right)$ determine the normalized rate of decay of the corresponding oscillation in the sense that the oscillation amplitude decays to $1/e$ in $1/(2\pi \xi)$ cycles \cite{kundur1994power}. For safe operation of the power system, the system must maintain a minimum damping ratio above $0.03\sim 0.05$ \cite{pal2006robust,chung2004generation}.

\subsection{Optimal Power Flow for Small-Signal Stability}
Under time-varying load and renewable generation, the damping performance also changes as the system equilibrium moves. When the system damping degrades too much, a small disturbance would create a large oscillation or even destabilize the system. Therefore, it is necessary to exploit the flexibility from both generation and demand sides to redirect the system to a safer equilibrium. The task of determining the optimal coordinated response of generation and demand to enhance system damping is therefore formulated as the following optimization problem:
	\vspace{-0.1cm}
	\begin{subequations}\label{WDROPF}
	\begin{equation}\label{WDROPF_obj}
	\underset{    
		V\in \mathbb{C}^{n_b}
	}{\text{max}} \xi_{\text{min}}(J(V)), \qquad \qquad 
		\vspace{-0.2cm}
	\end{equation}
	\begin{empheq}[left =  \mathrm{s.t.}\empheqlbrace\,]{align}
	 & \underline{P}_{g}	\leq \text{Re}\{ [C_g V]Y_g^* V^* \} \leq \overline{P}_{g} , \label{optcon3}\\
	    & \underline{Q}_{g}	\leq \text{Im}\{ [C_g V]Y_g^* V^* \} \leq \overline{Q}_{g} , \label{optcon4}\\
	    & \underline{P}_{l}	\leq -\text{Re}\{ [C_l V]Y_l^* V^* \} \leq \overline{P}_{l}, \label{optcon5}\\
	    & \underline{Q}_{l} \leq	-\text{Im}\{ [C_l V]Y_l^* V^* \} \leq \overline{Q}_{l}, \label{optcon6}\\
	    & \underline{V} \leq \abs{V} \leq \overline{V}, \label{optcon7} \\
	    & {\underline{S}_f \leq S_f(V) \leq \overline{S}_f,}  \label{optcon8}
	\end{empheq}
\end{subequations}
where $\xi_{\text{min}}(J(V))$ is the minimum damping ratio of the linearized system $J(V)$. {We choose this objective function because the method is designed to guide the corrective demand response and generator re-dispatch when the system's small-signal stability is at risk. Furthermore, considering other objective functions, such as generation cost and power loss, would bring no fundamental change to the algorithm.} Inequality constraints (\ref{optcon3}) and (\ref{optcon4}) reflect active and reactive power flexibility from the generators. Similarly, constraints (\ref{optcon5}) and (\ref{optcon6}) account for available flexibility from the load side. Constraints (\ref{optcon7}) and (\ref{optcon8}) set the security limits for bus voltage magnitudes and apparent power flow on transmission lines. For simplicity, we also write the problem in a compact form as
\begin{equation}\label{opt}
\underset{{x}}{\mathrm{max} }
\ \underset{i}{\mathrm{min} } \ \xi_i(J(x)), \ \text{s.t. }{h}\left({x}\right)\le 0,
\end{equation}
where $\xi_i = {-\alpha_i}/{\sqrt{\alpha_i^2+\beta_i^2}}$ and $\lambda_i = \alpha_i + j\beta_i$. {In constrast to other SSSC-OPF formulation, e.g. \cite{li2016sqp}, our problem formulation (\ref{WDROPF}) only involves the system complex voltages as explicit decision variables, making it consistent and compatible with the conventional OPF problem. The proposed SSSC-OPF problem formulation has the same scaling with the system size as the conventional OPF, which makes it possible to incorporate flexible loads and apply to large-scale systems.}

\subsection{Sequential Nonlinear Optimization Algorithm}

The difficulty in solving the above optimization problem largely comes from the non-smoothness of the objective function. Note that each eigenvalue is generically a smooth and differentiable function of the decision variable $x$. Therefore, in the neighborhood of an operation point $\hat{x}$, {the following linear program serves as a piece-wise linear approximation of the minimum damping ratio:}
\begin{equation}\label{piecelinear}
    f(x) ={\text{max } \gamma}\ \text{s.t. } {\xi }_i\left(J\left(\hat{x}\right)\right)+{\left.\frac{\partial {\xi }_i}{\partial x}\right|}_{\hat{x}}\left(x-\hat{x}\right) \geq \gamma ,\ \forall \ i,
\end{equation}
where the derivative of the damping ratio with respect to $x$ is given by
\begin{equation}\label{deriv1}
  \left.  \frac{\partial \xi_i}{\partial x} \right|_{\hat{x}} = \frac{\text{Im}(\lambda_i)}{\abs{\lambda_i}^3} \text{Im} \left( {\lambda}^*_i \left. \frac{\partial \lambda_i}{\partial x}\right|_{\hat{x}}\right).
\end{equation}
Here,
\vspace{-0.1cm}
\begin{equation}\label{deriv2}
    \begin{aligned}
 &   \left. \frac{\partial \lambda_i}{\partial x}\right|_{\hat{x}} =
 \frac{\partial}{\partial x}
 \Bigg(
 \frac{
 v_i^H
        \begin{pmatrix}
  A(x) & B(x) \\
C(x) & D(x) 
    \end{pmatrix}
      u_i}{{v}^H_i{u}_i}
      \Bigg) \Bigg|_{\hat{x}},
    \end{aligned}
\end{equation}
where $u_i=(u'_i,u''_i)$ and $v_i=(v'_i,v''_i)$ with ${u}'_i$ and ${v}'_i$ being the left and right eigenvectors of $J\left(\hat{x}\right)$ corresponding to ${\lambda }_i$, and ${u}''_i=-D(\hat{x})^{-1}C(\hat{x}){u}'_i$, ${v}''_i=-D(\hat{x})^{-T}B(\hat{x})^T{v}'_i$. {Details on the calculation of the derivative on the r.h.s.\ of (\ref{deriv2}) are included in the technical notes of our GitHub repository\footnotemark[\value{footnote}].} It is largely due to the introduction of analytic derivative (\ref{deriv1}) along with the piece-wise linear approximation (\ref{piecelinear}) that enables efficient optimization. Admittedly, the approximation (\ref{piecelinear}) is only valid in a small neighborhood of $\hat{x}$. In the optimization process, we maintain a ``trust region'' for such approximation, defined as the $l_{\infty }$-norm ball of radius $\tau $ centered at $\hat{x}$, and solve the following nonlinear program (NLP) in this region:
\begin{equation}\label{GrindEQ__4_}
\vspace{-0.0cm}
\underset{{x},\gamma}{\mathrm{max}}\ \gamma, \ 
 \mathrm{s.t.} \left\{
	\begin{aligned}
&{h}\left({x}\right)\le 0,\\
& {\xi }_i\left(J\left(\hat{x}\right)\right)+{\left.\frac{\partial {\xi }_i}{\partial x}\right|}_{\hat{x}}\left(x-\hat{x}\right) \geq \gamma ,\ \forall \ i, \\
& \norm{x-\hat{x}}_{\infty}<\tau,
\end{aligned}\right.
\end{equation}
where the radius $\tau $ is adaptively adjusted in the algorithm.

\begin{figure}[t]
	\includegraphics[width=3.6in]{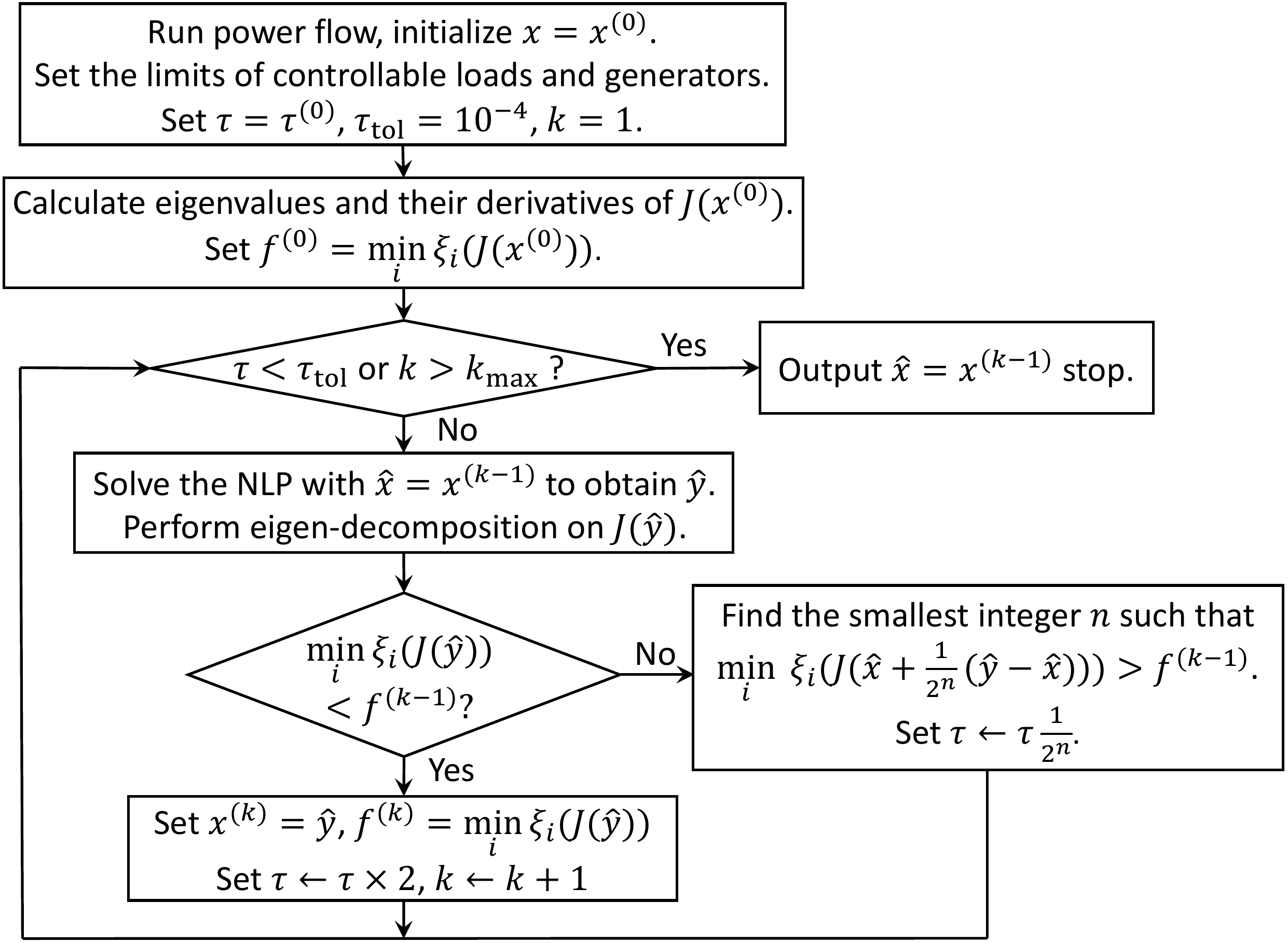}
	\caption{Flowchart of the SNLP algorithm for the transmission-level optimization. A Matlab implementation of this algorithm is available on our GitHub repository \cite{GitHubref}.} \label{SNLP}
	\vspace{-0.5cm}
\end{figure}

The flowchart of a basic sequential nonlinear program (SNLP) algorithm is shown in Fig. \ref{SNLP}. The major steps are illustrated as follows. Step 1: Initialize $x=x^{(0)}$ by running the power flow at {the initial} operation point. Specify the upper and lower limits of the controllable loads. Set $\tau ={\tau }^{(0)}$, $\tau_{tol}={10}^{-4}$, and $k=1$. Step 2: Compute all eigenvalues and eigenvectors of $J(x^{(0)})$ and calculate the derivatives of the damping ratios using \eqref{deriv1}. Set $f^{(0)}={\mathop{\mathrm{min}}_{i} {\xi}_i(J(x^{(0)}))}$. Step 3: If $\tau <\tau_{tol}$ or $k>k_{max}$, output $\hat{x}=x^{(k-1)}$ and stop. Otherwise, solve the NLP problem \eqref{GrindEQ__4_} with $\hat{x}=x^{(k-1)}$ using an interior point method (IPM) to obtain a solution $\hat{y}$. Perform eigen-decomposition on $J(\hat{y})$ to obtain all the eigen-information. Step 4: If ${\mathop{\mathrm{min}}_{i} {\xi }_i(J(\hat{y}))\ }>f^{(k-1)}$, i.e., the objective value has improved over the previous iteration, set $x^{(k)}=\hat{y}$, calculate the derivatives of the damping ratios using \eqref{deriv1}, $f^{(k)}={\mathop{\mathrm{min}}_{i} {\xi }_i(J(\hat{y}))\ }$, $\tau \leftarrow \tau \times $$2$, $k\leftarrow k+1$, and go to Step 3; otherwise, go to Step 5. Step 5: Find the least integer $n$ such that ${\mathop{\mathrm{max}}_{i} {\xi }_i\mathrm{(}J\mathrm{(}\hat{x}\mathrm{+}\frac{\mathrm{1}}{{\mathrm{2}}^n}\mathrm{(}\hat{y}\mathrm{-}\hat{x}\mathrm{)))}\mathrm{}\mathrm{>}f^{(k-1)}}$. Set $\tau \mathrm{\leftarrow }\tau \frac{\mathrm{1}}{{\mathrm{2}}^n}$. Go to Step 3.

Once the SNLP algorithm above finds a locally optimal solution $\hat{x}=(\hat{\theta},\hat{\mathcal{V}})$ of problem \eqref{WDROPF}, the generator setpoints for the exciter voltages and mechanical power are given by
\begin{equation}
    \hat{E}_f =  \text{Im}\{[\text{e}^{j(\frac{\pi}{2}-{\delta})}](C_g+jX_d Y_g)\hat{V} \},
\end{equation}
\begin{equation}
    \hat{P}_m= \text{Re}\{[C_g\hat{V}]Y_g^*\hat{V}^*\},
\end{equation}
respectively. The load power targets are determined through
\begin{equation}
  \hat{P}_l+j\hat{Q}_l = [C_l\hat{V}]Y_l^*\hat{V}^*.
\end{equation}
{Note that the proposed SNLP method does not rely on gradient sampling as used in \cite{li2016sqp}. Calculating the gradients at all sample points requires many eigen-decompositions which are computationally expensive for large systems. The SNLP algorithm only requires one eigen-decomposition in each iteration. {In addition, each iteration of the SNLP algorithm generates an improved feasible solution that can be implemented immediately.}}

\section{Distribution-level Control}
\label{sec:Dcon}
The basic task of distribution-level control is to distribute the power targets received from transmission-level optimization among all the buildings inside the distribution system, so that the distribution system can track the upper-level power targets without violating the network's voltage and capacity constraints. Moreover, the stability-oriented application considered in this article also puts a demanding requirement on the response speed of the distribution-level control. 
The control algorithm has to be simple and straightforward with negligible computational time. Note that the distribution network is usually an unbalanced three-phase radial system, and each building may be only connected to a single phase of the system. The controller first distributes the power target at the PCC among the three phases and then assign power targets to buildings sitting on different phases. 

Considering one of the distribution networks, the DL\_ctrl receives the aggregate power targets $\hat{P}_0$ and $\hat{Q}_0$ at the PCC. It is assumed that the DL\_ctrl has the phase-wise real-time power measurement $P_{0,\phi}(t)$ and $Q_{0,\phi}(t)$ at the PCC, where $\phi \in \{a,b,c\}$ is the phase and $P_{0}(t)=\sum_{\phi}P_{0,\phi}(t)$ and  $Q_{0}(t)=\sum_{\phi}Q_{0,\phi}(t)$ are the aggregate measurements.
The DL\_ctrl is a discrete-time system with sampling time $t_s=10$ms. The power targets at time $t$ sent to the $i$th building connected to phase $\phi$ are
\begin{equation}\label{contrl1}
   \hat{P}_{i,\phi}(t) =  \hat{P}_{i,\phi}(t-1) + k_I w^p_{i,\phi}(t) \left(\hat{P}_0-P_0(t-1)\right),
\end{equation}
\begin{equation}\label{contrl2}
   \hat{Q}_{i,\phi}(t) =  \hat{Q}_{i,\phi}(t-1) + k_I w^q_{i,\phi}(t) \left(\hat{Q}_0-Q_0(t-1)\right),
\end{equation}
where $k_I$ is a controller parameter resembling the integral parameter in Proportional Integral (PI) control. Parameter $ w^p_{i,\phi}(t)$ and $w^q_{i,\phi}(t)$ are respectively the active and reactive participation factors of the load at bus $i$ and phase $\phi$. {The participation factors are non-negative tunable parameters that reflect the maximum amount of power in the buildings available for participation in the demand response program.} They are voluntarily determined by the building occupants with incentives from some market mechanism.

Let $\mathcal{N}$ denote the set of all nodes in the distribution network and $\mathcal{N}_j$ be the subset of nodes that are adjacent to node $j$. To consider the voltage and line capacity constraints in the distribution network, we only need to interrupt the update of participation factors when a constraint violation is happening. Let $\mathcal{N}$ denote the set of all nodes in the distribution network and $\mathcal{N}_j$ is the subset of nodes that is adjacent to node $j$. The underlying principle stems from the DistFlow model \cite{baran1989network,chen2017robust,shi2015real} for the power flow of a radial distribution network:
\begin{subequations}\label{distflow}
\begin{equation}
    P_{ij,\phi} - r_{ij,\phi} \cdot \frac{P_{ij,\phi}^2+Q_{ij,\phi}^2}{U_{i,\phi}} = P_{j,\phi} + \sum_{k\in \mathcal{N}_{j}} P_{jk,\phi},
\end{equation}
		\begin{equation}
		    Q_{ij,\phi} - x_{ij,\phi} \cdot \frac{P_{ij,\phi}^2+Q_{ij,\phi}^2}{U_{i,\phi}} = Q_{j,\phi} + \sum_{k\in \mathcal{N}_{j}} Q_{jk,\phi},
		\end{equation}
		\begin{align}
		     &   U_{i,\phi}-U_{j,\phi} = 2\cdot (r_{ij,\phi}P_{ij,\phi}+x_{ij,\phi}Q_{ij,\phi}) \nonumber\\ 
 & \quad \quad \quad \quad  \quad \quad \quad \quad  -(r_{ij,\phi}^2+x_{ij,\phi}^2)\frac{P_{ij,\phi}^2+Q_{ij,\phi}^2}{U_{i,\phi}},
		\end{align}
\end{subequations}
where $P_{ij,\phi}$ and $Q_{ij,\phi}$ are the active and reactive power flow across the line from bus $i$ to bus $j$ with impedance $r_{ij,\phi}+jx_{ij,\phi}$ for phase $\phi$. 
The square of the voltage magnitude of bus $i$ in phase $\phi$ is denoted by $U_{i,\phi}$. It is apparent from the DistFlow equation (\ref{distflow}) that 1) the load power at any bus only affect its upstream line flows; 2) the load power at any bus affect both its upstream and down stream bus voltages on the same feeder. Therefore, we can adopt the following strategies to avoid constraint violation:
\begin{enumerate}
    \item If $P_{ij,\phi}^2(t)+Q_{ij,\phi}^2(t)>\overline{S}_{ij,\phi}^2-\epsilon_{s}$ and $dP_{0,\phi}(t)>0$ ($dQ_{0,\phi}(t)>0$), set $w^p_{k,\phi}(t)=0$ ($w^q_{k,\phi}(t)=0$) for all load $k$ connected to phase $\phi$ downstream of line $i$-$j$.
    
    \item If $U_{i,\phi}>\overline{U}_{i,\phi}-\epsilon_v$ and $dP_{0,\phi}(t)>0$ ($dQ_{0,\phi}(t)>0$), set $w^p_{k,\phi}(t)=0$ ($w^q_{k,\phi}(t)=0$) for all load $k$ connected to phase $\phi$ on the same feeder as bus $i$.
    
    \item If $U_{i,\phi}<\underline{U}_{i,\phi}+\epsilon_v$ and $dP_{0,\phi}(t)<0$ ($dQ_{0,\phi}(t)<0$), set $w^p_{k,\phi}(t)=0$ ($w^q_{k,\phi}(t)=0$) for all load $k$ connected to phase $\phi$ on the same feeder as bus $i$.
    
    \item {The participation factors are restored to the original values once the corresponding conditions that forced them to switch off are no longer met.}
\end{enumerate}

The $\epsilon_s$ and $\epsilon_v$ are two small constants to ensure voltage and capacity limits would not be crossed within one sample time of the controller. {The advantages of the proposed distribution-level control strategy are as follows:
\begin{enumerate}
\item it is a measurement-based feedback control {that only requires the knowledge of the topology}, with no need to acquire the detailed parameters (e.g., impedance) of the network;

\item the computation can be easily done in real time and is highly scalable to large systems;

\item it only requires the power measurements at the PCC and does not need access to the power measurement of each individual building, and hence protects the privacy of building occupants;

\item the only communications required are to broadcast the control error signal at the PCC and to update the switch-on/off status of the participation factors.

\end{enumerate}

The rest of this section is devoted to establishing a convergence guarantee for the DL\_ctrl. The active power $P_0$ and reactive power $Q_0$ at the PCC are functions of all the load power $\bm{P}=[P_{i,\phi}]_{i\in \mathcal{N},\phi \in \{a,b,c\}}$ and $\bm{Q}=[Q_{i,\phi}]_{i\in \mathcal{N},\phi \in \{a,b,c\}}$, i.e., $P_0 = \mathcal{P}(\bm{P},\bm{Q})$, $Q_0 = \mathcal{Q}(\bm{P},\bm{Q})$. The functions $\mathcal{P}(\cdot)$ and $\mathcal{Q}(\cdot)$ are implicitly given by the power flow equation (\ref{distflow}). {Similarly, we define $\bm{\hat{P}}=[\hat{P}_{i,\phi}]_{i\in \mathcal{N},\phi \in \{a,b,c\}}$ and $\bm{\hat{Q}}=[\hat{Q}_{i,\phi}]_{i\in \mathcal{N},\phi \in \{a,b,c\}}$}. If the network is assumed to be lossless, i.e., $r_{ij,\phi}=x_{ij,\phi}=0$ for all $i$ and $j$ (as assumed, e.g., by LinDistFlow model \cite{baran1989optimal}), one would have $\mathcal{P}(\bm{P},\bm{Q}) = \sum_{i,\phi} P_{i,\phi}$ and $\mathcal{Q}(\bm{P},\bm{Q}) = \sum_{i,\phi} Q_{i,\phi}$. However, in general, $\mathcal{P}(\cdot)$ and $\mathcal{Q}(\cdot)$ are nonlinear, continuously differentiable functions of $\bm{P}$ and $\bm{Q}$, and the active and reactive power flow are coupled with each other. For two different configurations of loads $(\bm{P},\bm{Q})$ and $(\bm{P'},\bm{Q'})$, applying the mean-value theorem yields
\begin{equation}\label{diffP}
\begin{aligned}
    \mathcal{P}&(\bm{P},\bm{Q}) - \mathcal{P}(\bm{P'},\bm{Q'}) \\
    &= \sum_{i,\phi} \alpha_{i,\phi}^{pp} (P_{i,\phi}-P'_{i,\phi}) + \sum_{i,\phi} \alpha_{i,\phi}^{pq} (Q_{i,\phi}-Q'_{i,\phi}),
    \end{aligned}
\end{equation}
\begin{equation}\label{diffQ}
\begin{aligned}
    \mathcal{Q}&(\bm{P},\bm{Q}) - \mathcal{Q}(\bm{P'},\bm{Q'}) \\
    &= \sum_{i,\phi} \alpha_{i,\phi}^{qp} (P_{i,\phi}-P'_{i,\phi}) + \sum_{i,\phi} \alpha_{i,\phi}^{qq} (Q_{i,\phi}-Q'_{i,\phi})
    \end{aligned}
\end{equation}
for some scalar $ \alpha_{i,\phi}^{pp}$, $ \alpha_{i,\phi}^{pq}$, $ \alpha_{i,\phi}^{qp}$, and $ \alpha_{i,\phi}^{qq}$ that depend on $(\bm{P},\bm{Q})$ and $(\bm{P'},\bm{Q'})$. For a lossless network, $\alpha_{i,\phi}^{pp}=\alpha_{i,\phi}^{qq}=1$ and $\alpha_{i,\phi}^{pq}=\alpha_{i,\phi}^{qp}=0$. Therefore, for a practical lossy network, we assume that there is a small scalar $\epsilon$ such that $\abs{\alpha_{i,\phi}^{pp}-1}\leq \epsilon$, $\abs{\alpha_{i,\phi}^{qq}-1}\leq \epsilon$, $\abs{\alpha_{i,\phi}^{pq}}\leq \epsilon$, and $\abs{\alpha_{i,\phi}^{qp}}\leq \epsilon$ for all $i$ and $\phi$. 

From the power flow equation (\ref{distflow}), we can observe that, for the distribution network, the PCC power $\mathcal{P}(\cdot)$ and $\mathcal{Q}(\cdot)$ are monotonically increasing with respect to each $P_{i,\phi}$ and $Q_{i,\phi}$. Therefore, the scalar $ \alpha_{i,\phi}^{pp}$, $ \alpha_{i,\phi}^{pq}$, $ \alpha_{i,\phi}^{qp}$, and $ \alpha_{i,\phi}^{qq}$ can all be chosen to be non-negative. In addition, we assume that the response of each load follows a first-order transfer function:
\begin{equation}\label{loadp}
P_{i,\phi}(t) = P_{i,\phi}(t-1) + \frac{t_s}{T_{i,\phi}}(\hat{P}_{i,\phi}(t)-P_{i,\phi}(t-1)),
\end{equation}
\begin{equation}\label{loadq}
Q_{i,\phi}(t) = Q_{i,\phi}(t-1) + \frac{t_s}{T_{i,\phi}}(\hat{Q}_{i,\phi}(t)-Q_{i,\phi}(t-1)),
\end{equation}
where $T_{i,\phi}$ is the time constant of the load at bus $i$ and phase $\phi$. Therefore, the whole closed-loop system is described by (\ref{contrl1}), (\ref{contrl2}), (\ref{loadp}), and (\ref{loadq}) with $P_0(t) = \mathcal{P}(\bm{P}(t),\bm{Q}(t))$ and $Q_0(t) = \mathcal{Q}(\bm{P}(t),\bm{Q}(t))$. 

We make the following assumption.
\begin{assumption}\label{assp1}(i) $\epsilon<\frac{1}{4}$, (ii) $\frac{t_s}{T_{i,\phi}}<2$ for $i\in \mathcal{N}$ and $\phi \in \{a,b,c\}$, and {(iii) $0<\sum_{i,\phi}w^p_{i,\phi}(t), \sum_{i,\phi} w^q_{i,\phi}(t)\leq 1 $}.
\end{assumption}
Assumption 1(i) means that the active and reactive power losses of the network are less than 25\% of the aggregate active and reactive power at the PCC, respectively, which is reasonable in practice since the actual loss is usually less than a few percent in real networks. Assumption 1(ii) means that the sampling time of the controller is less than twice of the time constant of any load response. In our implementation, $t_s=10$ ms which is far smaller than the time constant of any load. Assumption 1(iii) means that the transfer capacity and flexibility of the distribution network is adequate in the sense that at least one load participation factor is not switched to zero by the strategies to avoid constraint violation. 

Now, we establish the convergence of the control strategy given by (\ref{contrl1}) and (\ref{contrl2}) in the following theorem.
\begin{theorem}
For the control strategy given by (15) and (16) with sufficiently small parameter $k_I>0$ and under Assumption 1, we have $\lim_{t \to \infty}P_0(t) = \hat{P}_0$ and $\lim_{t \to \infty}Q_0(t) = \hat{Q}_0$.
\end{theorem}
\begin{IEEEproof}
By defining the following variables:
\begin{equation}
    \Delta P_{i,\phi} (t) = \hat{P}_{i,\phi}(t) - P_{i,\phi} (t),
\end{equation}
\begin{equation}
    \Delta Q_{i,\phi} (t) = \hat{Q}_{i,\phi}(t) - Q_{i,\phi} (t),
\end{equation}
\begin{equation}
    \Delta P_{0} (t) = \hat{P}_{0} - \mathcal{P}(\bm{P}(t),\bm{Q}(t)),
\end{equation}
\begin{equation}
    \Delta Q_{0} (t) = \hat{Q}_{0} - \mathcal{Q}(\bm{P}(t),\bm{Q}(t)),
\end{equation}
\begin{equation}
    M(t) = \mathcal{P}(\bm{\hat{P}}(t),\bm{\hat{Q}}(t)) -\mathcal{P}(\bm{P}(t),\bm{Q}(t)),
\end{equation}
\begin{equation}
    N(t) = \mathcal{Q}(\bm{\hat{P}}(t),\bm{\hat{Q}}(t)) -\mathcal{Q}(\bm{P}(t),\bm{Q}(t)),
\end{equation}
\begin{equation}
\begin{aligned}
    \beta^{pp} = \sum_{i,\phi} \alpha_{i,\phi}^{pp} w^p_{i,\phi}, \ \   \beta^{pq} = \sum_{i,\phi} \alpha_{i,\phi}^{pq} w^p_{i,\phi},\\
    \end{aligned}
\end{equation}
\begin{equation}
    \beta^{qp} = \sum_{i,\phi} \alpha_{i,\phi}^{qp} w^q_{i,\phi}, \ \ \beta^{qq} = \sum_{i,\phi} \alpha_{i,\phi}^{qq} w^q_{i,\phi},
\end{equation}
the dynamical system given by equations (\ref{contrl1}), (\ref{contrl2}), (\ref{loadp}), and (\ref{loadq}) can be equivalently written as
equations (\ref{eqdyno1})-(\ref{eqdyno4}) shown at the top of the next page.

\newpage
\begin{strip}
\begin{equation}\label{eqdyno1}
  \Delta P_{i,\phi}(t) = (1-\frac{t_s}{T_{i,\phi}}) \Delta P_{i,\phi}(t-1) + k_I w^p_{i,\phi}(1-\frac{t_s}{T_{i,\phi}})  \Delta P_{0}(t-1)  + k_I w^p_{i,\phi}(1-\frac{t_s}{T_{i,\phi}}) M(t-1),
\end{equation}
\begin{equation}\label{eqdyno2}
  \Delta Q_{i,\phi}(t) = (1-\frac{t_s}{T_{i,\phi}}) \Delta Q_{i,\phi}(t-1) + k_I w^q_{i,\phi}(1-\frac{t_s}{T_{i,\phi}})  \Delta Q_{0}(t-1)  + k_I w^q_{i,\phi}(1-\frac{t_s}{T_{i,\phi}}) N(t-1),
\end{equation}
\begin{equation}\label{eqdyno3}
    \Delta P_{0}(t) = (1-k_I \beta^{pp})\Delta P_{0}(t-1) -k_I \beta^{pp}M(t-1)-k_I \beta^{pq}\big(\Delta Q_{0}(t-1)+N(t-1)\big),
\end{equation}
\begin{equation}\label{eqdyno4}
    \Delta Q_{0}(t) = (1-k_I \beta^{qq}\Delta Q_{0}(t-1) -k_I \beta^{qq}N(t-1)-k_I \beta^{qp}\big(\Delta P_{0}(t-1)+M(t-1)\big).
\end{equation}
\hspace{0cm}----------------------------------------------------------------------------------------------------------------------------------------------------------
\end{strip}

It is straightforward to verify that $\abs{\beta^{pp}-1} \leq \epsilon$, $\abs{\beta^{qq}-1} \leq \epsilon$, $\abs{\beta^{pq}} \leq \epsilon$, and $\abs{\beta^{qp}} \leq \epsilon$. In addition, by the Cauchy-Schwarz inequality, we have
\begin{equation}\label{bdM}
  \frac{1}{6n}  M^2(t) \leq (1+\epsilon)^2\sum_{i,\phi} \Delta P^2_{i,\phi}(t) + \epsilon^2\sum_{i,\phi} \Delta Q^2_{i,\phi}(t),
\end{equation}
\begin{equation}\label{bdN}
  \frac{1}{6n}  N(t)^2 \leq  \epsilon^2\sum_{i,\phi} \Delta P^2_{i,\phi}(t) + (1+\epsilon)^2\sum_{i,\phi} \Delta Q^2_{i,\phi}(t),
\end{equation}
where $n$ is the number of nodes in the distribution network.

We define the function
\begin{equation}\label{lyap}
    V(t) = \sum_{i,\phi} \Delta P^2_{i,\phi} (t) + \sum_{i,\phi} \Delta Q^2_{i,\phi} (t) + \Delta P^2_{0} (t) + \Delta Q^2_{0} (t).
\end{equation}
We will show that $V(t)$ is a Lyapunov function of the dynamical system. Substituting equations (\ref{eqdyno1}), (\ref{eqdyno2}), (\ref{eqdyno3}), and (\ref{eqdyno4}) into equation (\ref{lyap})
, and taking advantage of equations (\ref{bdM}) and (\ref{bdN}), for any scalar constant $\gamma >0$, we have
\begin{equation}\label{Vdiff2}
    \begin{aligned}
   & V(t)-V(t-1) \\
    & \leq \sum_{i,\phi} \big( -\frac{t_s}{T_{i,\phi}}(2-\frac{t_s}{T_{i,\phi}})\big) \Delta P^2_{i,\phi} (t-1)\\
    & + \sum_{i,\phi} \big( -\frac{t_s}{T_{i,\phi}}(2-\frac{t_s}{T_{i,\phi}})\big) \Delta Q^2_{i,\phi} (t-1)\\
    &+ k_I \sum_{i,\phi} (2+6n\gamma(2\epsilon^2+2\epsilon+1))  \Delta P^2_{i,\phi} (t-1)\\
    & +k_I \sum_{i,\phi} (2+6n\gamma(2\epsilon^2+2\epsilon+1))  \Delta Q^2_{i,\phi} (t-1)\\
    & + k_I 
X(t-1)^T
    \begin{bmatrix}
    A & B \\ B^T & -\gamma \bm{I}
    \end{bmatrix}
    X(t-1)
    + o(k_I),
    \end{aligned}
\end{equation}
where $X(t-1)^T=[\Delta P_{0} (t-1),\Delta Q_{0} (t-1),M(t-1),$ $N(t-1)]^T$, and 
\begin{equation}
    A = 
    \begin{bmatrix}
    1-2\beta^{pp}  &  -\beta^{pq}-\beta^{qp}\\
    -\beta^{pq}-\beta^{qp} & 1-2\beta^{qq}
    \end{bmatrix},
\end{equation}
\begin{equation}
    B = 
    \begin{bmatrix}
    -\beta^{pp}  &  -\beta^{pq}\\
    -\beta^{qp} & -\beta^{qq}
    \end{bmatrix}.
\end{equation}
Here, all the second order terms in $k_I$ are represented by $o(k_I)$. By the Gershgorin circle theorem, the matrix $A$ is negative definite. It can be verified that, if we set the scalar $\gamma > \frac{(2\epsilon+1)^2}{1-4\epsilon}$, the matrix $ \begin{bmatrix}
    A & B \\ B^T & -\gamma {I}
    \end{bmatrix}$ is then negative definite. For sufficiently small {value of $k_I$ depending on parameters $n$, $t_s$, $\epsilon$, and $T_{i,\phi}$, $\forall i\in \mathcal{N},\ \phi \in \{a,b,c\}$}, the sum of the first four right-hand-side terms of equation (\ref{Vdiff2}) is negative for non-zero $X(t-1)$. With such small $k_I>0$, $V(t)-V(t-1)<0$ for any non-zero $X(t-1)$, i.e., $V(t)$ is a Lyapunov function of the dynamical system given by (\ref{eqdyno1}), (\ref{eqdyno2}), (\ref{eqdyno3}), and (\ref{eqdyno4}). Thus, as $t \to \infty$, we have $V(t) \to 0$, which implies $P_0(t)\to \hat{P}_0$ and $Q_0(t)\to \hat{Q}_0$.
\end{IEEEproof}
}

\section{Building-level Control}
\label{sec:Bcon}
Each controllable building is assumed to possess one energy management controller (EMC). Once the EMC receives the power target from the DL\_ctrl, it schedules the controllable devices in the building to track the power target as closely as possible. The buildings have two types of controllable loads. Type-I loads are those whose on/off operations can be flexible within some operational constraints. Examples include a washer, dryer, dishwasher, etc. For type-II loads, the power consumption can be continuously adjusted. Examples include an electric vehicle whose charging rate can be changed as long as the vehicle is completely charged within a specified time interval, and an air conditioner whose power consumption can be changed if the resulting thermal comfort is within acceptable range specified by the consumer. We denote the sets of type-I and type-II loads by $\mathbb{K}_t$ and  $\mathbb{Z}_t$. In practice, the sets $\mathbb{K}$ and $\mathbb{Z}$ can be time-dependent because of human intervention and device constraints. {Following \cite{chen2014distributed}, a mixed integer linear programming (MILP) model is used to coordinate type-I and type-II devices in each building. }
Given the target $\hat{P}(t)$ at time $t$, each EMC decides the on/off status of loads in set $\mathbb{K}_t$ and adjusts the power consumption of appliances in set  $\mathbb{Z}_t$ by solving the following MILP problem:
\begin{subequations}\label{BLoptmodel}
\begin{align}
    \max_{\bm{x}(t), \bm{f}(t)} \sum _{i \in \mathbb{K}_t} x_{i}(t) \cdot f_{i} + \sum _{j \in \mathbb{Z}_t } v_{j}(t),
\end{align}
\begin{empheq}[left =  \mathrm{s.t.}\empheqlbrace\,]{align}
 &  \hat{q}(t) +\sum _{i \in \mathbb{K}_{t}}x_{i}(t) \cdot f_{i}  +\sum _{j \in \mathbb{Z}_t}v_{j}(t)  \le \hat{P}(t), \\
&  \xi _{j}(t) \cdot v_{j}^{\min}(t) \le v_{j}(t) \le \xi _{j}(t) \cdot v_{j}^{\max}(t),\\
&  x_{i}(t), \xi_{j}(t) \in \left\{0,1\right\}.
\end{empheq}
\end{subequations}
In this problem, $\hat{q}(t)$ is the total power consumption of the uncontrollable devices, $x_{i}(t)$ is the binary variable used to represent the on/off status of the $i$th type-I load, $f_{i}$ is the power consumed by the $i$th type-I load, $v_{j}(t)$ is the power consumed by the $j$th type-II load, $\xi _{j}(t) $ is the binary variable used to represent on/off status of the $j$th type-II load, and $v_{j}^{\min}(t)$ and $v_{j}^{\max}(t) $ are the power limits for the $j$th type-II load.

\begin{figure}[t]
\centering
	\includegraphics[width=2.2in]{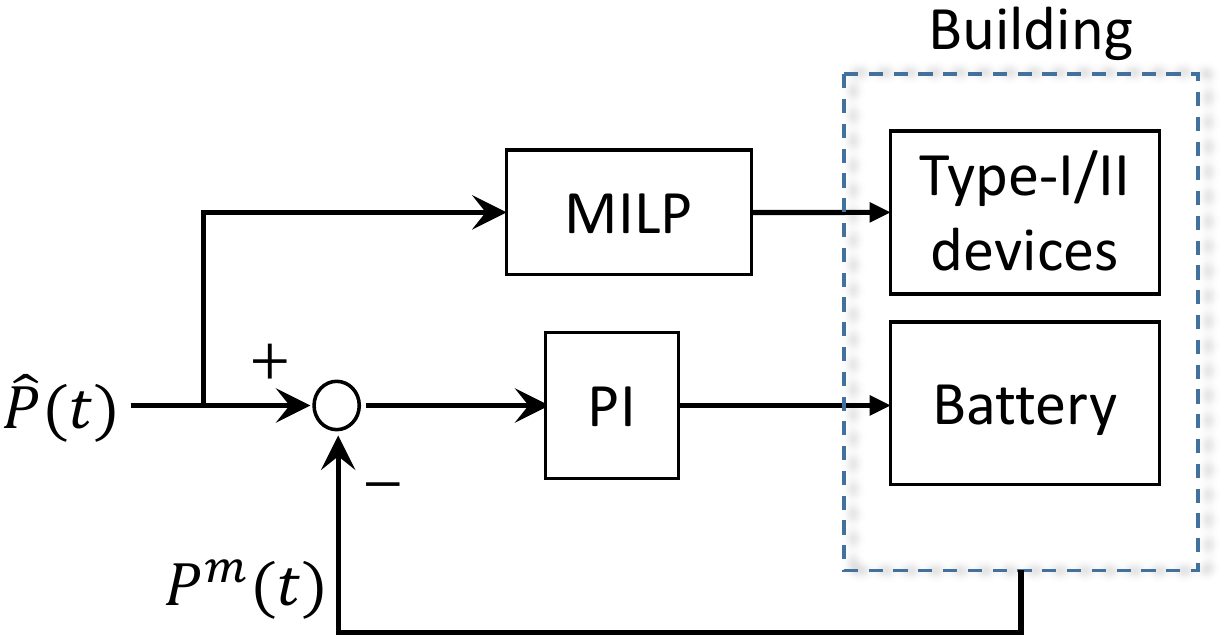}
	\caption{Building-level controller. Type-I/II devices are scheduled by the MILP solver and a fast-response battery is controlled by a PI controller in parallel.}
	\label{BL_ctrl}
\end{figure}

In practice, merely relying on the above optimization scheduling does not provide satisfactory steady-state and dynamical performance for two reasons. First, there could be characterization errors for some devices for which the actual power consumption deviates substantially from the nominal values. Second, some devices may respond much faster than others and the optimization model (\ref{BLoptmodel}) cannot consider the response discrepancy and further exploit those faster devices. In our implementation, we separate a fast type-II device (normally a battery) from the rest of the controllable devices and directly apply a PI control to this type-II device in parallel with the optimization (\ref{BLoptmodel}), as shown in Fig. \ref{BL_ctrl}. In this way, the steady-state control error can be eliminated and the dynamical performance can be significantly improved. {In addition, we assume that the reactive power tracking is provided by one or several converters in the building.}

\section{Integrated Real-time HiL Test}
\label{sec:hil}

\subsection{{Implementation}}

\begin{figure*}[h!] 
\centering
	\includegraphics[width=5.6in]{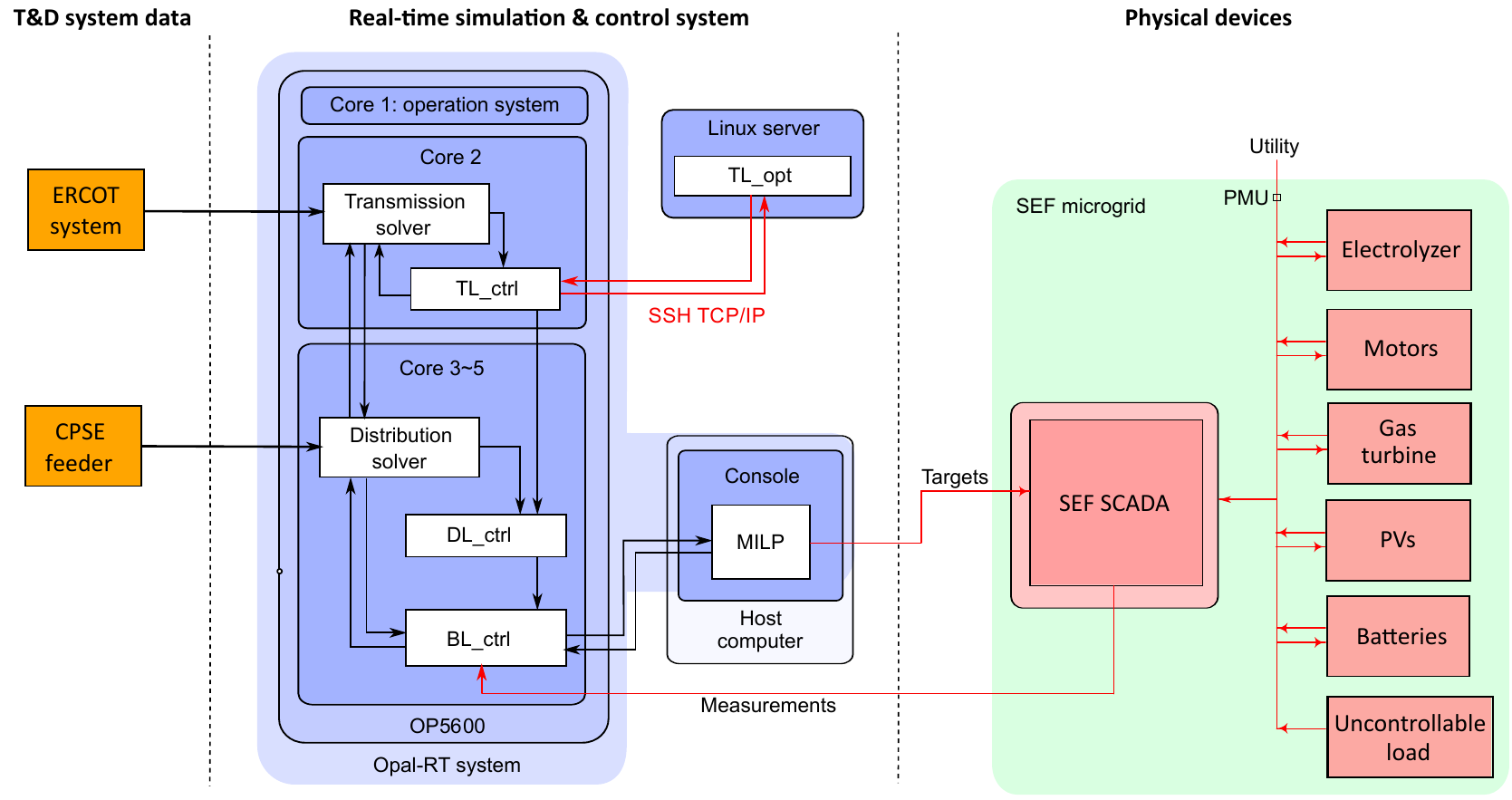}
	\caption{{Configuration of the integrated real-time HiL testbed.}}
	\label{hiltest}
\end{figure*}

{
To implement the whole control architecture in an environment that is as realistic as possible, we perform tests in an integrated HiL testbed designed using Opal-RT's real-time simulation system. The overall configuration of the testbed is shown in Fig. \ref{hiltest}. The Opal-RT system consists of two computers. One is called \emph{Target} (OP5600), a customized multi-processor Linux computer designed to run the simulation and communicate with external devices in real time. The other is called \emph{Host} and communicates asynchronously with the real-time process on Target, providing an interface between the user and Target. The testbed further integrates large-scale transmission and distribution network models, a real microgrid, as well as external computing resources. The whole system has the following key features:
\begin{enumerate}
    \item \emph{Co-simulation of large-scale transmission and distribution systems in real time.} The OP5600 real-time simulator is configured to run phasor-domain simulations on the ERCOT transmission system (7459 buses, 8926 lines, and 429 generators) and one of the feeders in the CPS Energy (CPSE) distribution system (3275 nodes and 4615 loads) using 10 ms time steps. We randomly select a subset of loads in the CPSE feeder as controllable buildings, so that they account for approximately 20\% of the total active power and are almost evenly distributed across the three phases. The two systems are modeled in detail using ePHASORSIM. Simulations of these systems along with all communication processes are run on four computer cores of Target, with the transmission system on one core and the distribution system and communication on the other three. The CPSE feeder is assumed to be connected to the ERCOT system at a selected bus, and it is considered one of the distribution networks in the ERCOT system. To model the coupling of the two networks, dynamical information is interchanged among the computer cores every time step. For the transmission system simulation, the distribution network model is treated as a lumped dynamic load whose impedance is adjusted every time step according to the power consumption at the distribution substation. Conversely, for the distribution system simulation, the transmission system is replaced by its Thevenin equivalent whose impedance and voltage are updated every time step to deliver the required power to the distribution network and maintain the voltage at the substation bus. The whole interconnected model can run in real time on OP5600. 
      \item \emph{Integration of real-time simulation and optimization.} In the proposed control architecture, two optimization algorithms need to be executed (one at the transmission level and the other at the building level). As these algorithms are iterative, they are not guaranteed to finish within each simulation time step. Therefore, they have to be configured outside of Target and communicate asynchronously with the real-time processes. The transmission-level optimization TL\_opt is run in an external high-performance Linux server that interfaces with Target through TCP/IP. The parallel computing capability of this server ensures that the transmission-level optimization can be performed effectively and efficiently. The building-level optimization BL\_OPT is run on Host, which is directly connected to Target via an Ethernet connection.
      \item \emph{Building-level hardware in the loop.} To account for the actual responses of real-world smart buildings or microgrids and their impact on the distribution and transmission-level dynamics, the Opal-RT system is installed in the microgrid at Stone Edge Farm (SEF) and connected to the SEF control system through the Modbus communication protocol. For the testing purpose, the SEF microgrid is regarded as one building in the CPSE feeder. The PMU measurement of SEF PCC power is fed into the simulation as the real-time load at one selected bus in the distribution system, and the building-level control commands are sent through Modbus to control 15 real devices at SEF, including 2 motors, 1 gas turbine, 1 electrolyzer, 2 PV panels, and 9 different types of storages. {Among these devices, the two motors belong to type I, as they can only be controlled on or off. All others are type-II devices whose output power can be continuously adjusted within a certain range.}
\end{enumerate}}
In this setup, we have a detailed model of one distribution network (the CPSE feeder) and the real response from one building (the SEF microgrid). For the other distribution networks connected to the ERCOT system and the other controllable buildings in the CPSE feeder, we build simulation models to represent their responses in the transmission and distribution systems, respectively. Our strategy is to first run a building-level test and identify a transfer function model for the SEF microgrid and then use this transfer function to represent the other controllable buildings in the CPSE feeder. Next, we run a distribution-level simulation and identify a transfer function model for the aggregate power fo the CPSE feeder, and then use this model to represent the responses of the other distribution networks in the ERCOT. After the transfer function models are implemented, we can run the whole real-time HiL testbed with the three levels of the system and control integrated together.

\subsection{Test Results}

\begin{figure}[t]
\centering
	\includegraphics[width=3.0in]{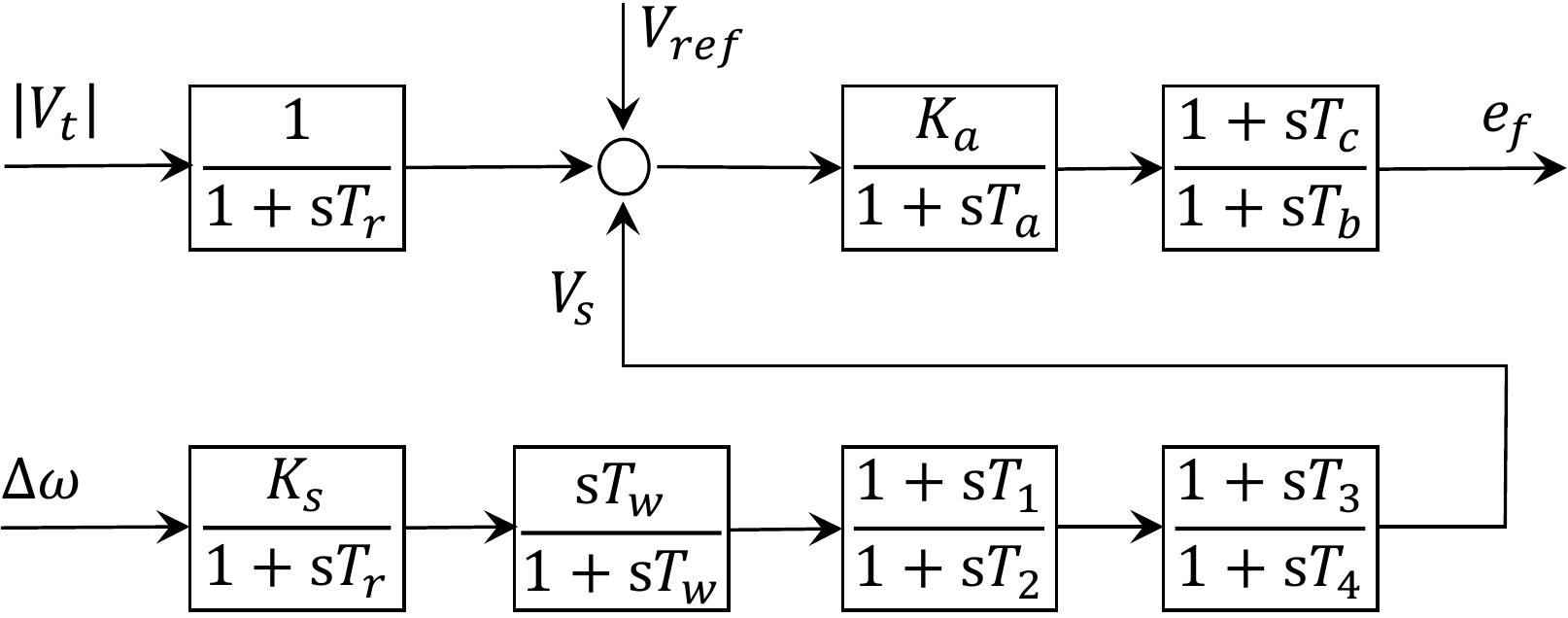}
	\caption{{Transfer function model for exciter and PSS.}}
	\label{exitorPSS}
\end{figure}

\begin{figure}[t]
		\centering
		\includegraphics[width=3.4in]{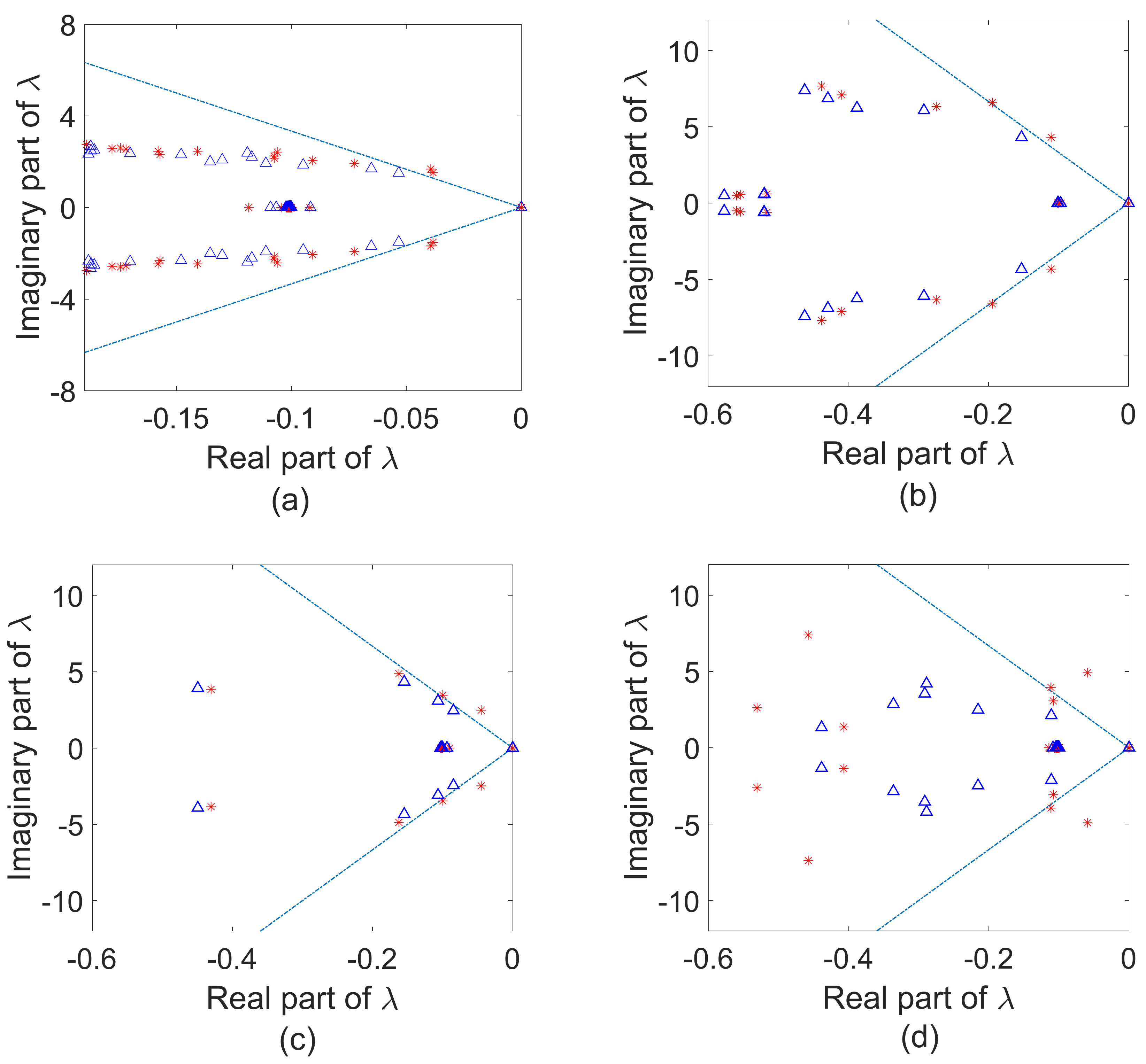}
		\caption{{Comparison of the system eigenvalues before ($\star$) and after ({$\boldsymbol{\bigtriangleup}$}) optimization on benchmark systems. The results correspond to: (a) the ERCOT system, (b) the 10-machine New England System, (c) the 16-machine NETS/NYPS system, and (d) the 50-machine Iowa system. The blue dot-dashed lines indicate the 3\% damping ratio. All plots are based on detailed models with exciters and PSSs.}} \label{4meigen}
	\end{figure}

We run the whole simulation starting from the ERCOT nominal power flow solution. 
The Opal-RT simulator establishes communication to a Linux computer ({i9-7980 CPU and 125GB RAM}) in which the transmission-level optimization solver is called. For online operation, we set $k_{max}=1$ and run the SNLP algorithm in parallel with $12$ different $\tau_0$ evenly distributed in $(0,1)$. Within 1 min, the Linux server solves the $12$ parallel problems and sends the best solution back to the Opal-RT simulator. The transmission controller then sends the setpoints for generator mechanical power and exciter voltages to generation units, and also active and reactive power targets to the distribution networks. Fig. \ref{4meigen}(a) shows a comparison between the system eigenvalues corresponding to the system equilibrium before and after optimization. The minimum damping ratio is improved from 0.0235 to 0.0354 (the improvement is 50.6\%), under the assumption that there is $\pm 10\%$ flexibility from all distribution networks and all the generators can be adjusted within its capacity limits. We also provide results from TL\_opt on three well-known benchmark systems in Fig. \ref{4meigen}(b), (c) and (d). It shows that the minimal damping ratios have all been significantly improved. {The mean and standard deviation of the solver time for one iteration of the transmission optimization over 100 different load levels (with the load factors evenly distributed in $[0,1]$) are listed in Table \ref{solvertime} for all four test systems.} Each iteration of the algorithm generates a feasible solution that can be readily implemented and thus satisfies the real-time requirement of the control scheme. Although the TL\_opt algorithm is developed with the simplified system model, the performance is validated on a detailed model in which exciters and PSSs are represented by the transfer functions shown in Fig. \ref{exitorPSS}.

\renewcommand{\arraystretch}{1.2}
\begin{table}[]
{
\caption{{TL\_opt solver time (mean$\pm$std).}}
\label{solvertime}
\begin{tabular}{|l|c|c|c|c|}
\hline
{\bf System}          & 10-machine & 16-machine & 50-machine & ERCOT \\ \hline
{\bf Solver time} (s) & 0.86$\pm$0.53       & 0.98$\pm$0.24      & 1.27$\pm$0.56       & 49.3$\pm$8.3 \\ \hline
\end{tabular}
}
\end{table}

\begin{figure}[t]
\includegraphics[width=3.4in]{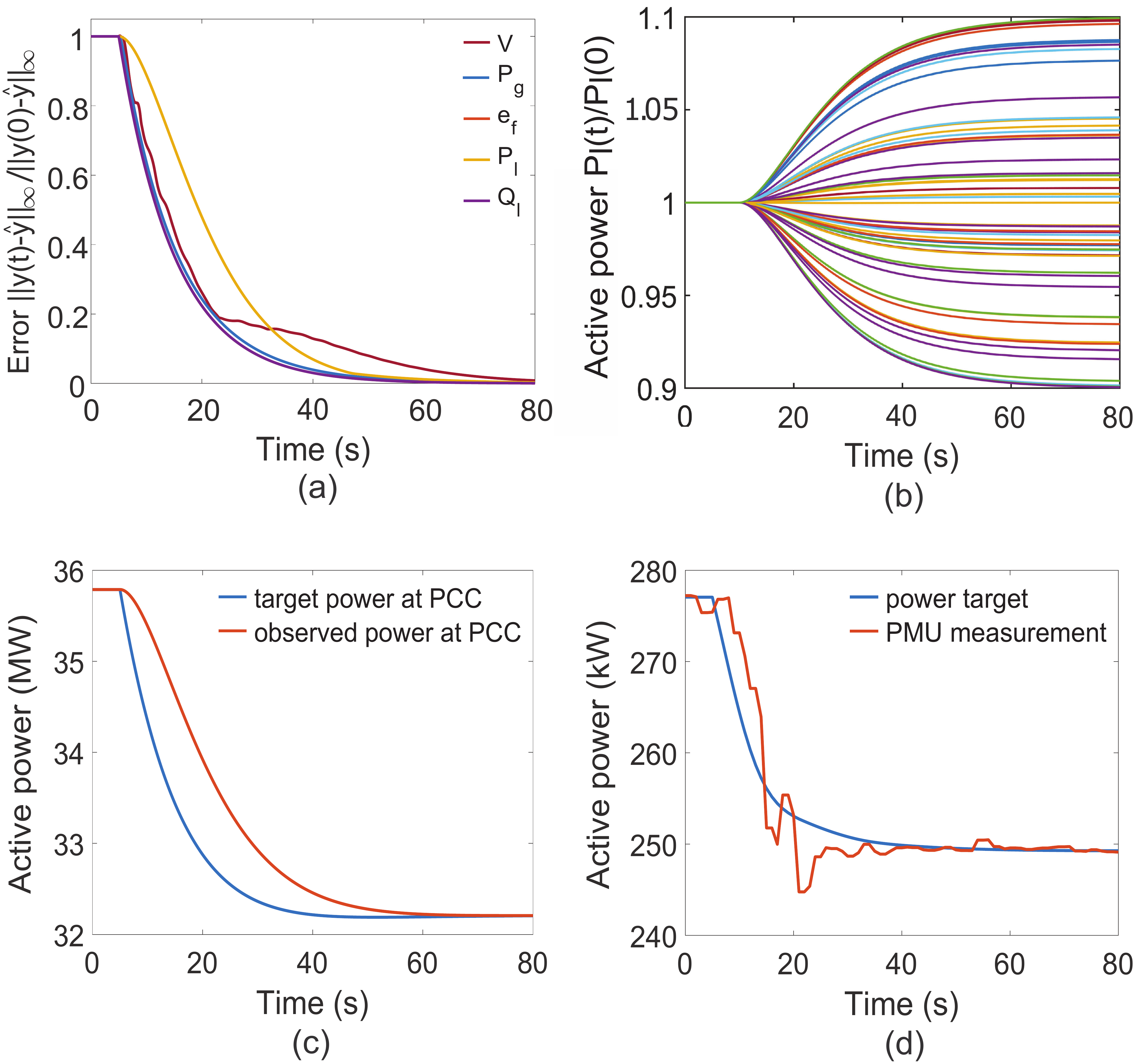}
\caption{System response during the implementation of the ERCOT optimization solution: (a) normalized control errors at the transmission level, (b) normalized active power of all distribution networks, (c) target and observed active power of the CPSE feeder, (d) target and observed active power in the SEF microgrid.} \label{control_error}
\vspace{0cm}
\end{figure}

After the optimization is completed, the transmission-level controller sends control commands to all the distribution networks ($P_l,Q_l$) and generation units ($P_m,E_f$), so that the system complex voltage states would start to move towards the target power flow solution, i.e., a new equilibrium with better synchronization stability. Figure \ref{control_error} (a) shows the normalized control error $\norm{y(t)-\hat{y}}_{\infty}/\norm{y(0)-\hat{y}}_{\infty}$ during the control implementation process where $y$ represents $P_m,E_f,P_l,Q_l$ and the system complex voltages $V$. We see that the transmission system indeed evolves to the desired power flow states. Figure \ref{control_error} (b) illustrates the change of active power at PCCs of all the distribution networks in the ERCOT system. In order to improve the synchronization of the transmission system, some distribution networks reduce their load while others counter-intuitively increase their load. This shows the highly nonlinear nature of the stability-oriented load assignment problem. 

\begin{figure}[t]
 \centering
	\includegraphics[width=3.4in]{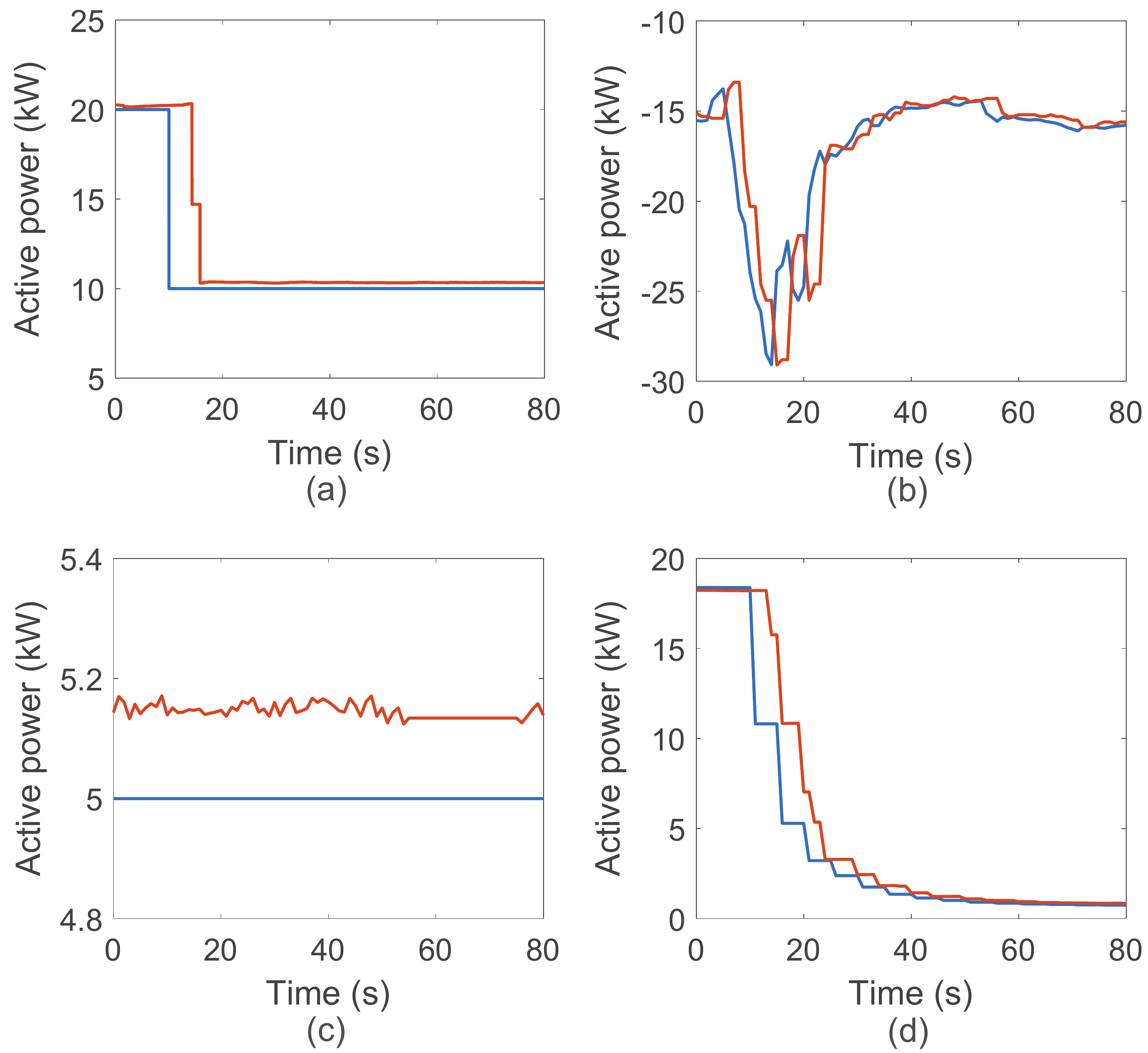}
	\caption{SEF device power targets (blue) and measurements (red) during the implementation of transmission-level control. (a) electrolyzer, (b) Tesla battery, (c) SolarEdge battery, and (d) Aquion battery.} \label{devicepower}
\end{figure}

The response of the detailed model of the CPSE feeder is shown in Fig. \ref{control_error}(c), and the response of the real building (the SEF microgrid) is shown in Fig. \ref{control_error}(d). Furthermore, some of the device-level responses in SEF are also shown in Fig. \ref{devicepower}, including the electrolyzer, as well as the Tesla, SolarEdge, and Aquion batteries. {The mean and standard deviation of the solver time for the building-level optimization over 100 different power target values (evenly distributed in the feasible range of the SEF microgrid) are 0.28 s and 0.17 s, respectively. This satisfies the requirement to update the control command every 1.0 s.} In the building-level control implementation, the Tesla battery is the device directly controlled by the PI controller in Fig. \ref{BL_ctrl}. It is apparent in Fig. \ref{devicepower} that the actual responses of devices are subject to different degrees of time delays and characterization errors, whose adverse effects are largely neutralized by the fast-response Tesla battery with local PI control. Without the PI control loop in Fig. \ref{BL_ctrl}, the overall building response would be much slower and subject to steady-state errors. 

\begin{figure*}[t]
\centering
\includegraphics[width=6.4in]{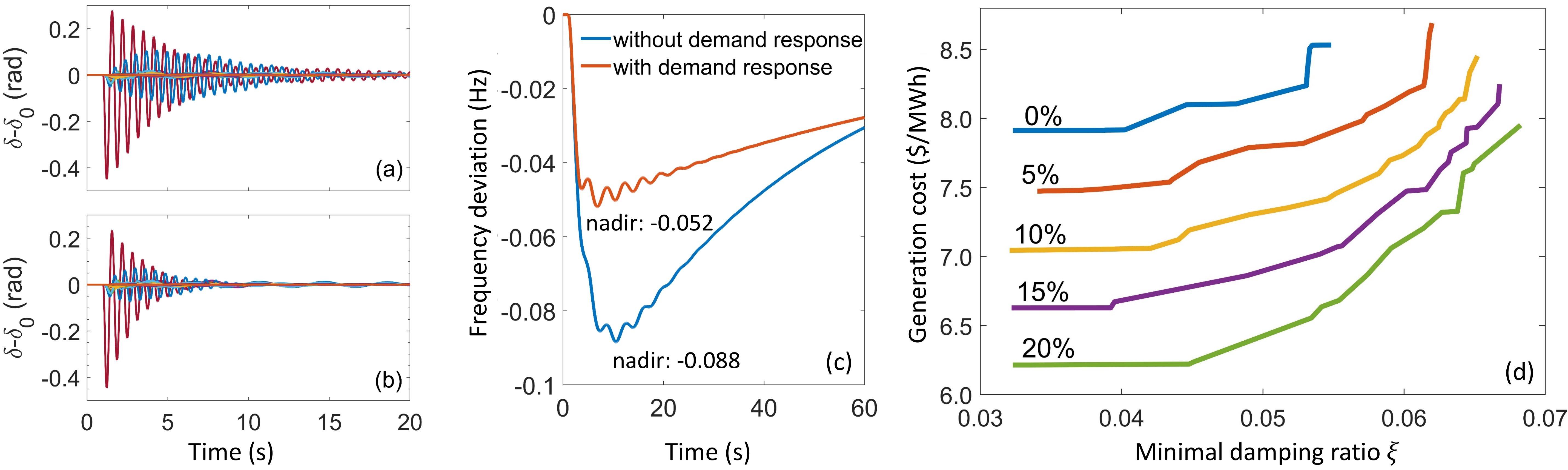}
	\caption{Rotor angle deviation during disturbance (a) before and (b) after optimization in the ERCOT system. (c) System frequency as a function of time during the disturbance. {(d) Optimal generation cost (normalized by the total load demand) as a function of the optimal minimal damping ratio in the ERCOT system under different levels (0\%, 5\%, 10\%, 15\%, 20\%) of flexibility from the loads. Here, we set $k_{max}=5$ in the SNLP algorithm.}} \label{disturbcompare}
\end{figure*}

To show the effects of the control on the system damping performance, we apply the same disturbance to the system operating at the original and optimized power flow solutions. The disturbance we created is a three-phase line-to-ground fault at bus 6007. The fault happens after 1.0 s of the simulation and then all cleared by the protection relay in 100 ms. Figure \ref{disturbcompare}(a)-(b) compares the generator rotor angle deviations from the corresponding equilibria after the fault. We can see that, when operated at the optimized power flow, the disturbance is suppressed much quicker than when the system is operated at the original power flow.

From the frequency stability perspective, the transmission-level control in Fig. \ref{ctrldiagram} not only involves the conventional load frequency control but also invokes the frequency support from the demand side. To assess the effects of DSR on the system frequency stability, we compare the system frequency response with and without DSR when a sudden loss of 2300 MW generation happens to the system. It is shown in Fig. \ref{disturbcompare}(c) that, with the help of demand-side frequency support, the frequency nadir under the specified disturbance has been improved from -0.088 Hz to -0.052 Hz. This also means that the system with demand-side response can withhold heavier contingency under the same operation standard.

{To further explore the trade-off between generation cost and stability level, we run the transmission-level optimization in the ERCOT system with the objective function $f(x) = c(x) - \rho \cdot \min_i \ \xi_i(J(x))$, where the term $c(x)$ is the total generation cost and the second term is the minimal damping ratio. The generation cost of each unit is uniformly sampled between \$0/MWh and \$20/MWh. The non-negative parameter $\rho$ reflects the trade-off between the two objective functions. For each given level of flexibility from the load demand, we run the optimization for a wide range of the parameter $\rho$ and plot the optimized generation cost as a function of the corresponding minimal damping ratio. The results are shown in Fig. 10(d), where each curve represents a different level of flexibility from the load demand. We can see that the curves corresponding to higher load flexibility are located at much lower positions, meaning that the load flexibility significantly contributes not only to the reduction of generation cost but also to the improvement of system stability.}




\section{Discussion}
\label{sec:con}
In this article, we presented a hierarchical generation- and demand-coordinated control architecture. It enhances small-signal rotor angle stability and frequency stability of the transmission system by real-time feedback control of not only the generating units but also the distribution networks, which is realized through optimal scheduling of controllable devices in buildings. Our real-time HiL test with real system data and actual building-level devices shows that the proposed control scheme is computationally capable of online operation and significantly improves the damping performance and frequency recovery ability under faults and sudden generation loss. {A major lesson from the HiL tests is that the controllable devices are usually far from ideal in the following aspects. First, many devices and the associated communication channels have significant time delays, which vary substantially from device to device. Control parameters must be designed respecting different time delays. Second, many devices have large characterization errors, meaning that their steady-state power consumption may differ a lot from their nominal values. This poses significant challenges for optimal scheduling of devices. Therefore, feedback controllers must be implemented in addition to the optimal scheduling algorithm to eliminate steady-state errors. Third, we may have a mixture of continuously and discretely adjustable devices with widely varying response times, requiring the optimization algorithm and control strategy to work for a hybrid system with multiple time scales. Extensions of this work that warrant further research include decentralization of the transmission and distribution control and development of a systematic approach to deal with complex human behavior in the building-level control.}

\appendices
\section*{Acknowledgment}
We thank CPS Energy for providing the feeder circuit data, Stone Edge Farm for providing access to their microgrid, and Opal-RT and Heila Technologies for assistance with the HiL setup and testing.

\bibliographystyle{IEEEtran}

\long\def\/*#1*/{}

\/*

\begin{IEEEbiographynophoto}
{Chao Duan}
received the B.S. degree from Xi'an Jiaotong University, Xi'an, China, in 2012 and dual Ph.D. degrees from Xi'an Jiaotong University, Xi'an, China, and the University of Liverpool, Liverpool, U.K., in 2018, all in electrical engineering. In 2018 he joined the Department of Physics and Astronomy at Northwestern University, Evanston, IL, USA, first as a Postdoctoral Fellow and then as a Research Assistant Professor. His research interests include power system operation, network dynamics, and system control.
\end{IEEEbiographynophoto}

\begin{IEEEbiographynophoto}
{\bf Pratyush Chakraborty}  received B.E. degree in electrical engineering from Jadavpur University, Kolkata, India, in 2006; M.Tech. degree in electrical engineering from IIT Bombay, Mumbai, India, in 2011; M.S. and PhD. degrees in electrical and computer engineering from the University of Florida, Gainesville, USA in 2013 and 2016, respectively. From 2006 to 2009, he worked for Siemens Limited, India. From 2017 to 2020, he worked as Postdoctoral Researcher at three institutions in the USA: UC Berkeley, Northwestern University, and University of Utah. He is currently an Assistant Professor in the Department of Electrical and Electronics Engineering, BITS Pilani, Hyderabad, India. His research interests include game theory, optimization, and applications to cyber-physical-social systems.
\end{IEEEbiographynophoto}

\begin{IEEEbiographynophoto}
{\bf Takashi Nishikawa} received the Ph.D. degree in applied mathematics from the University of Maryland, College Park, MD, USA, in 2000 and his postdoctoral training at Arizona State University, Tempe, AZ, USA. After holding faculty positions at Southern Methodist University, Dallas, TX, USA, and at Clarkson University, Potsdam, NY, USA, he joined the faculty of Northwestern University, Evanston, IL, USA, in 2012, where he has been a Research Professor of Physics. His research focuses on mathematical and computational analysis of network dynamical systems. He is a fellow of the American Physical Society.
\end{IEEEbiographynophoto}

\begin{IEEEbiographynophoto}
{Adilson E. Motter}
received the Ph.D. degree from the University of Campinas, Campinas, Brazil, in 2002 and his postdoctoral training from the
Max Planck Institute for the Physics of Complex Systems,
Dresden, Germany, and Los Alamos National Laboratory,
Los Alamos, NM, USA.  He joined the faculty of Northwestern University, Evanston, IL, USA, in 2006, where he has served as Chair Professor of Physics since 2011. His research is focused on the mathematical and computational analysis of network dynamical systems. He is a fellow of the American Physical Society, the Network Science Society, and the American Association for the Advancement of Science.
\end{IEEEbiographynophoto}

*/

\end{document}